\documentclass[prl,twocolumn,showpacs]{revtex4}
\usepackage{graphicx}
\font\got=eufm10 scaled \magstep1 \font\gotscr=eufm7 scaled
\newfam\gotfam
\textfont\gotfam=\got \scriptfont\gotfam=\gotscr
\scriptscriptfont\gotfam=\gotscrscr
\def\got{\fam\gotfam}

\font\Bbb=msbm10 scaled \magstep1 \font\Bbbscr=msbm7 scaled
\newfam\Bbbfam
\textfont\Bbbfam=\Bbb \scriptfont\Bbbfam=\Bbbscr
\scriptscriptfont\Bbbfam=\Bbbscrscr
\def\Bbb{\Bbbfam}

\font\Cal=msbm10 scaled \magstep1 \font\Calscr=msbm7 scaled
\newfam\Calfam
\textfont\Calfam=\Cal \scriptfont\Calfam=\Calscr
\scriptscriptfont\Calfam=\Calscrscr
\def\Cal{\fam\Calfam}
\def\beq{\begin{equation}}
\def\eeq{\end{equation}}
\def\gappeq{\mathrel{\rlap {\raise.5ex\hbox{$>$}}
{\lower.5ex\hbox{$\sim$}}}}

\def\lappeq{\mathrel{\rlap{\raise.5ex\hbox{$<$}}
{\lower.5ex\hbox{$\sim$}}}}

\newcommand{\lsim}{\raise.3ex\hbox{$<$\kern-.75em\lower1ex\hbox{$\sim$}}}
\newcommand{\gsim}{\raise.3ex\hbox{$>$\kern-.75em\lower1ex\hbox{$\sim$}}}
\begin{document}
\title{Transverse Momentum Dependence of Intercept  Parameter $\lambda$
       of Two-Pion (-Kaon) Correlation Functions in q-Bose Gas Model}

%\author{D.V.~Anchishkin$^{a}$,
%A.M.~Gavrilik$^{a}$, and S.Y.~Panitkin$^{b,}$}

\author{D.V.~Anchishkin}
\affiliation{Bogolyubov Institute for Theoretical Physics,
          03143 Kiev, Ukraine}
\email{anch,omgavr@bitp.kiev.ua}
\author{A.M.~Gavrilik}
\affiliation{Bogolyubov Institute for Theoretical Physics,
          03143 Kiev, Ukraine}
%\email{omgavr@bitp.kiev.ua}
\author{S.Y.~Panitkin}
 \affiliation{Brookhaven National Laboratory, Upton,
New York 11973, U.S.A. }
\email{panitkin@bnl.gov}

\date{\today}

\begin{abstract}
Within recently proposed approach aimed to effectively describe
the observed non-Bose type behavior of the intercept $\lambda$ of
two-particle correlation function $C(p,K)$ of identical pions or
kaons detected in heavy-ion collisions, the $q$-deformed
oscillators and $q$-Bose gas picture are employed.
For the intercept $\lambda$, connected with deformation
parameter $q$, the model predicts a fully specified dependence
of $\lambda$ on pair mean momentum ${\bf K}$.
The intercepts $\lambda_\pi$ and $\lambda_K$ for pions and kaons,
differing noticeably at small ${\bf K}$, should merge at ${\bf K}$
large enough, i.e., in the range $|{\bf K}|\ \ge \! 800$ MeV/c,
where the effect of resonance decays is negligible.
In this paper we confront, fixing $q$ appropriately, the predicted
dependence $\lambda_\pi=\lambda_\pi({\bf K})$ with recent results
from STAR/RHIC for $\pi^-\pi^-$ and $\pi^+\pi^+$ pairs, and find
nice agreement.
Using the same $q$, we also predict behavior of $\lambda$ for kaons.

\end{abstract}

\maketitle

%\narrowtext

%%%%%%%%%%%%%%%%%%%%%%%%%%%%%%%%%%%%%%%%%%%%%%%%%%%%%%%%%%%%%%%%%%%%%
%\section{Introduction}
%\label{sec1}
%%%%%%%%%%%%%%%%%%%%%%%%%%%%%%%%%%%%%%%%%%%%%%%%%%%%%%%%%%%%%%%%%%%%%

Two-particle correlations in momentum space can be used to extract
information about the space-time structure of the emitting sources
created in heavy ion collisions. The method exploits in an
essential way the quantum mechanical uncertainty relation between
coordinates and momenta, and thus any formal treatment of
two-particle correlations must be based on a quantum mechanical
description. For so-called "chaotic" sources where the two
particles are emitted independently the description can be based
on the {\it single-particle} Wigner density $S(x,K)$ of the
source (source function).

In standard quantum mechanical treatment the Bose-Einstein
correlations are due to symmetrization of the two-particle
(many-particle) wave function (suppose particles are emitted
independently) $ \psi_{\gamma_a \gamma_b}({\bf x}_a,{\bf x}_b,t) =
\frac{1}{\sqrt{2}} \big[ \psi_{\gamma_a}({\bf x}_a,t)\,
         \psi_{\gamma_b}({\bf x}_b,t) \,
 + \, e^{i \alpha} \psi_{\gamma_a}({\bf x}_b,t)\,
         \psi_{\gamma_b} ({\bf x}_a,t) \big]
$ with $\alpha =0$ ($\alpha =\pi $) for identical bosons
(fermions).
The indices $\gamma_a,\gamma_b$ of the 1-particle wave
functions label complete sets of 1-particle quantum numbers.
In the following, we consider two-particle correlations of
noninteracting spin zero identical bosons.
The correlation
function, with $P_1\left({\bf k}\right)$ and $P_2\left({\bf k}_a,
{\bf k}_b\right)$ being single- and two-particle probabilities to
detect particles with given momenta, is defined as
\begin{equation}
C({\bf k}_a,{\bf k}_b)= \frac{\displaystyle P_2\left({\bf k}_a,
{\bf k}_b\right) } {\displaystyle P_1\left({\bf k}_a\right) \,
P_1\left({\bf k}_b\right) } . \nonumber
\end{equation}
\noindent
In the absence of final state interactions (FSI, see \cite{AHR}),
for chaotic source, the correlation function can be expressed
as follows     \cite{AGI}:
\begin{eqnarray}                                     \label{2}
C({\bf k}_a,{\bf k}_b)  = 1\, + \, \cos{\alpha} \, \frac{ \left|
\int  d^4 x \, e^{i  p\cdot x } S(x,K) \right| ^2 } {
 \int  d^4 x \, S\big( x,k_a \big) \,
 \int  d^4 y \, S\big( y,k_b \big)
}
%\nonumber
\end{eqnarray}
with the 4-momenta $K=\frac{1}{2} (k_a+k_b)$ as pair mean momentum
and $ p=k_a-k_b $ as relative momentum.
%%%%%%%%     \debug{Variable X is not defined.}
The source function $S(x,K)$
%(single-particle Wigner density)
is defined by the single-particle states $\psi_\gamma (x)$ at
freeze-out time and the source density matrix
$\rho_{\gamma \gamma '}$ as, e.g., in \cite{AGI}.
Obviously, from (\ref{2}) at zero relative momentum
${\bf k}_a={\bf k}_b$  one gets
$C({\bf k}_a,{\bf k}_a) = 1 + \cos\alpha \equiv 1+\lambda$
and since for bosons $\alpha=0$ it follows that
$C({\bf k}_a,{\bf k}_a) = 2$, i.e., $\lambda=1\, .$
To fit experimental data, the correlation
function of {\it identical bosons} is usually presented as
$C(p,K) = 1\, + \,  \lambda \, f(p,K)$, with
$f(p,K)$ commonly taken as Gaussian so that $f({\bf p}=0,K)=1$.
From the very first experiments it was deduced that $\lambda $
is lesser than one, the typical experimental values being
$\lambda=$ 0.4 - 0.9~.
%
%>From the very first experiments it was deduced that the factor
%$\cos\alpha$ in (\ref{2}) is lesser than one. The correlation
%function of {\it identical bosons} is usually presented as
%$C(p,K) = 1\, + \,  \lambda \, f(p,K)$ with $\lambda $
%drawn from a fit to
%a data, with typical experimental values $\lambda=$ 0.4 - 0.9 and
%$f(p,K)$ commonly taken as Gaussian so that $f({\bf p}=0,K)=1$.
%
The second term in (\ref{2}) is obviously due to
quantum-mechanical interference, and deviation of
$\lambda $ from unity manifests weakening of the interference
effects which can occur due to different reasons - influence of
long lived resonances, coherent emission, etc.
%One of existing explanations of deviation of the parameter
%$\lambda$ from unity suggests that (at least one of) the particles
%e.g., pions, originates from long-lived resonances. Then, with
%today registrating capabilities these correlations cannot be
%resolved because of ${|\bf p|} \le $5--10$ MeV/c$.
%\noindent
%-------------------------------------------------------------------------\\
%{\bf What about correlations of kaons from long lived resonances ?
%What kind of resonances can give the same contribution as for pions ? }\\
%-------------------------------------------------------------------------
%{\bf  Idea :}

Let us explain the key idea of the model developed  in
\cite{AGI,AGI2} (named AGI-model in what follows) and further
exploited in this letter.
In two-boson correlations the deviation
of intercept $\lambda$ from unity, besides the contribution
due to effects from long-lived resonances, can also be caused by
the averaged softening of quantum-statistical effects in the
peculiar short-lived many-particle systems formed in relativistic
heavy ion collisions.
In such small system, the symmetrization angle $\alpha$ of
$\psi_{\gamma_a \gamma_b}({\bf x}_a,{\bf x}_b,t)$ can be distorted
by an additional phase due to a nonhomogenuity of the system
at freeze-out times (strong radial and azimuthal flows).
These peculiarities can cause the effect analogous to Aharonov-Bohm one.
As result, a finite value of averaged symmetrization angle may appear:
$\overline{\alpha } > 0$ for bosons and $ \overline{\alpha } < \pi$
for fermions.

Now, trying to explain experimental data with formula (\ref{2}),
it is natural to relate the parameter $\lambda $ with the
averaged angle $\overline{\alpha }$ to get {\it the reduction
factor $\lambda$} by means of $\cos{\overline{\alpha}}$.
That is, the deviation of intercept $\lambda$ from
unity is viewed to be due to fluctuations of symmetrization
angle $\alpha$, i.e.
\begin{equation}
\lambda=\cos{\overline{\alpha }}\, .
\label{4}
\end{equation}
Notice that slow bosons (pions, kaons) will experience bigger
fluctuations (deviations) of symmetrization angle $\alpha$ than
the particles with high velocities in fireball frame.
That is, the deviation of intercept $\lambda$ from unity for slow bosons
should be more sizable than for the fast ones.

To implement our key idea we exploit quantum field theory with
$q$-deformed commutation relations (qDCR) and the techniques
of $q$-boson statistics  (see \cite{VZ} and refs. therein)
which reflects a partial suppression of the quantum
statistical effects.
In \cite{tsallis94,arik99} it was argued that the algebra of qDCR is
connected, {\it for real $q$ only}, with the so-called nonextensive
statistics introduced by Tsallis     \cite{tsallis88}.
This type of generalized statistics has already found numerous
applications in diverse branches of modern physics (see
\cite{tsallis99} for refs.). In particular, nonextensive
statistics was applied to the problems of high-energy
nuclear collisions (\cite{UWW00} and refs. therein).
However, the techniques of $q$-boson statistics based on qDCR
allows the use of complex values as well as the real values for
the deformation parameter $q$, depending on the choice of
algebraic realization of qDCR. The physical reasons for usage of
qDCR, and subsequent interpretation of $q$, essentially differ
depending on the case of $q$ real or complex. Introducing deformed
statistics with $q$ real enables one to effectively account for
interaction effects by means of non-interacting ideal gas of
``modified" particles. On the other hand, the approach based on
qDCR provides the ability to model the effects involving the
Aharonov-Bohm like phase, intimately connected with symmetrization
properties of wave functions.

For the system of pions or kaons produced in heavy ion collisions,
we employ the ideal $q$-Bose gas picture. Physical meaning or
explanation of the origin of $q$-deformation in the considered
phenomenon sharply differs in the case of real deformation
parameter $q$ from the case when $q$ is a pure phase factor, as
will be seen in what follows.

The AGI-model exploits two different sets of qDCR. The first is
the multimode Biedenharn-Macfarlane (BM-type) $q$-oscillator
defined as \cite{BM}: $[N_j,b_j]=-b_j ,\ $ $[N_j,b^\dagger_j]=
b^\dagger_j ,\ $ $b_j b_j^\dagger-q^{-1} b_j^\dagger b_j=q^{N_j}
,$
where different modes ($i\ne j$) commute. Then, $b^\dagger_i
b_i=[N_i]_q$ (here the ``$q$-bracket'' means
$[r]_q=(q^r-q^{-r})/(q-q^{-1})$ ) so that $b^\dagger_i b_i = N_i$
is recovered in the ``classical'' (``no deformation'')
limit $q\to 1$ .
Below, for the BM-type $q$-oscillators it is meant that
\begin{equation}                                     \label{5}
q=\exp (i \theta)\ ,
          \ \ \ \ \ \ \ \ \ \ 0 \le \theta < \pi/2 \ .
\end{equation}

%\vspace{4mm}
%\noindent
The second multimode $q$-oscillator used in AGI-model is the set
of Arik-Coon (AC-type) $q$-oscillators, defined by the relations
\cite{AC}
\smallskip
$ [{\cal N},a]=-a, $ $ [{\cal N},a^\dagger]= a^\dagger, $ and $ a
a^\dagger-q a^\dagger a=1$ (subscript suppressed). Again, at
\mbox{$q\!\ne\!1$}, the bilinear $a^\dagger_i a_i$ does not
equal the number operator ${\cal N}_i$ (as is true for usual
bosonic oscillators, i.e., at $q=1$).
Instead, $a^\dagger_i a_i=[[{\cal N}_i]] $ where
now the notation $[[r]] \equiv (1-q^r)(1-q) , $ is used.
The $q$-bracket $[[\hat A]]$ for an operator $A$ is understood
as a formal series.  At $q\to 1$, from $[[\hat A]]$
one recovers $\hat A$.
In what follows we set
\beq                                               \label{6}
 -1\le q\le 1\ .
\eeq
For each such value of the {\it deformation parameter} $q$, the
$a^\dagger_i\ $ and  $\ a_i$ are mutual conjugates. Note that the
inverse of the relation $a^\dagger_i a_i=[[{\cal N}_i]] $ is given
by a formula expressing the operator ${\cal N}_i$ as a formal series
of creation/annihilation operators.

%%%%%%%%%%%%%%%%%%%%%%%%%%%% sec2 %%%%%%%%%%%%%%%%%%%%%%%%%%%%%%%%%%
%% \section{Single-particle distribution function}
%%%%%%%%%%%%%%%%%%%%%%%%%%%%%%%%%%%%%%%%%%%%%%%%%%%%%%%%%%%%%%%%%%%%

For a multi-pion (-kaon) system, viewed as ideal gas of
$q$-bosons the Hamiltonian is taken as
\begin{equation}
H=\sum_i{\omega_i { N}_i}                           \label{7}
\end{equation}

\vspace{-0.4cm} \noindent with $i$ labelling energy eigenvalues,
$\omega_i=\sqrt{m^2+{\bf k}_i^2}$, and $N_i$ defined as above.
This is unique truly noninteracting Hamiltonian with additive
spectrum \cite{VZ}. We assume discrete $3$-momenta of particles
(the system is in a box of volume $\sim L^3$). {}For the set of
AC-type $q$-oscillators, one takes ${\cal N}_i$ instead of $N_i$
in (\ref{7}).

Statistical properties are obtained by evaluating thermal averages
$ \langle A \rangle ={\rm Sp}(A\rho )/{\rm Sp}(\rho)\ $, $\rho =
{\rm e}^{-\beta H}$ , with the Hamiltonian (\ref{7}) and
$\beta=1/T$.

With $b^\dagger_i b_i=[N_i]_q$ and $q+q^{-1}=[2]_q=2\cos\theta$,
the $q$-deformed distribution function is obtained as  \cite{VZ,AG,AGI}
%
%\begin{equation}
%\langle b_i^\dagger b_i \rangle=\frac{{\rm e}^{\beta\omega_i}-1}
%{{\rm e}^{2\beta\omega_i}-2\cos(\theta){\rm e}^{\beta\omega_i}+1}
%\ .
%\label{10}
%\end{equation}
%
\begin{equation}                                      \label{8}
\langle b_i^\dagger b_i \rangle
        =\frac{1}{{\rm e}^{\beta\omega_i}-1+\delta_i}\
, \ \ \ \ \delta_i = 2~\frac{1-\cos\theta}
                            {\ 1-{\rm e}^{-\beta\omega_i}} \ .
\end{equation}
If $\theta\to 0$, it yields Bose-Einstein (BE) distribution.
%,as should ($q=1$ recovers usual bosonic commutation relations).
Note that the $q$-distribution function
%%%% \debug{I fixed label for exuation 8. PLease check if this is right}
(\ref{8}) is real.

The $q$-distribution (\ref{8}) deviates from the quantum
Bose-Einstein just in the ``right direction'' towards the
classical Boltzmann distribution, that
{\it reflects a decreasing of quantum statistical effects.}
For kaons, whose mass $m_K$ is bigger than $m_\pi$, analogous
curve should lie closer, than pion's one, to that of BE
distribution \cite{AGI2}.

In the case of AC-type $q$-bosons with real $q$ from (\ref{5}),
one arrives at the distribution function (cf.   \cite{VZ,AG,AGI}):
\beq \langle a_i^\dagger a_i \rangle=\frac{1}{{\rm
e}^{\beta\omega_i}-q}\ .
%\label{8}
\eeq
In the no-deformation limit $q\to 1$, this also reduces to the
Bose-Einstein distribution, since at $q=1$ we return to the
standard system of bosonic commutation relations.

The deviation from standard BE statistics is natural thing if one
considers the system of interacting particles versus that of
non-interacting particles (ideal gas).
For instance, natural type
of interaction is the hard-core repulsion of the particles that
assumes particle finite self-volume.
This type of interaction, as was shown in  \cite{anch92}, results in
the same kind (\ref{8}) of modified statistics.
At microscopical level, a finite self-volume
arising due to composite structure of particles results in
$q$-deformed commutation relations \cite{avancini95}  and
subsequently results in certain $q$-deformed statistics of the gas
of such particles.

%%%%%%%%%%%%%%%%%%%%%%%%%%%%%%% sec3 %%%%%%%%%%%%%%%%%%%%%%%%%%%%%%%
%% \section{ Intercept $\lambda$ }
%%%%%%%%%%%%%%%%%%%%%%%%%%%%%%%%%%%%%%%%%%%%%%%%%%%%%%%%%%%%%%%%%%%%

The two-particle distribution corresponding to the BM-type
$q$-oscillators is
\begin{equation}                                      \label{9}
\langle b_i^\dagger b_i^\dagger b_i b_i\rangle=\frac{2\cos\theta}
{{\rm e}^{2\beta\omega_i}-2\cos(2\theta){\rm e}^{\beta\omega_i}+1}
\ .
\end{equation}
{}From this and eq.~(\ref{7}), one obtains the intercept
$\tilde\lambda_i\equiv\lambda_i + 1= \langle b_i^\dagger
b_i^\dagger b_i b_i\rangle /(\langle b_i^\dagger b_i \rangle)^2$
of two-particle correlations (subscript omitted) as
\begin{equation}                                      \label{10}
%\tilde
\lambda=-1+ \frac{ 2\cos\theta\ (\cosh(\beta\omega)-\cos\theta)^2}
  {(\cosh(\beta\omega)-2\cos^2\theta+1)(\cosh(\beta\omega)-1)}.
\end{equation}
At $\beta\omega\to\infty$ (i.e., at low temperature and fixed
momenta or large momenta and fixed temperature)
the asymptotics of intercept is given merely by the deformation
angle $\theta$ (recall that $q=\exp({\rm i} \theta)$):
\begin{equation}                                     \label{11}
   %\hspace{1mm}
\lambda = \lambda^{\rm asymp} =
                       2\cos\theta -1 \hspace{6mm}
      (T\to 0 \ \ {\rm or} \ \  \vert{\bf K}\vert \to\infty) .
\end{equation}
{}From this and eq.~(\ref{4}) we have the (asymptotical) relation
$\cos\theta= \cos^2\frac{\overline{\alpha }}{2}$.
Note that, if the unique cause forcing the intercept
to be lesser than one is the decays of resonances
(the conventional viewpoint), all the curves would tend to
the value $\lambda =1$ in the large $|{\bf K}|$ limit.
In contrast, we predict a constant $\lambda <1$, as in (\ref{11}).

In the case of AC-type $q$-oscillators, the formula $ \langle
a_i^\dagger a_i^\dagger a_i a_i\rangle=
%\frac
({1+q}) {({\rm e}^{\beta\omega_i}-q)^{-1}({\rm
e}^{\beta\omega_i}-q^2)^{-1}}$ for two-particle distribution
combined with (\ref{8}) leads to  %yield for intercept the expression:
\beq {\lambda}=-1+\langle a^\dagger a^\dagger a a\rangle/
\langle a^\dagger a \rangle^2=
q-\frac{q~(1-q^2)}{{\rm e}^{\beta\omega}-q^2} .       \label{12}
\eeq
In this case, for $T\to 0\ $ or $\ |{\bf K}|\to\infty$
we have ${\lambda}^{\rm asymp}=q$.

\begin{figure}
%[t]
\begin{center}
\vspace{-0.01cm}
\includegraphics[width=9cm]{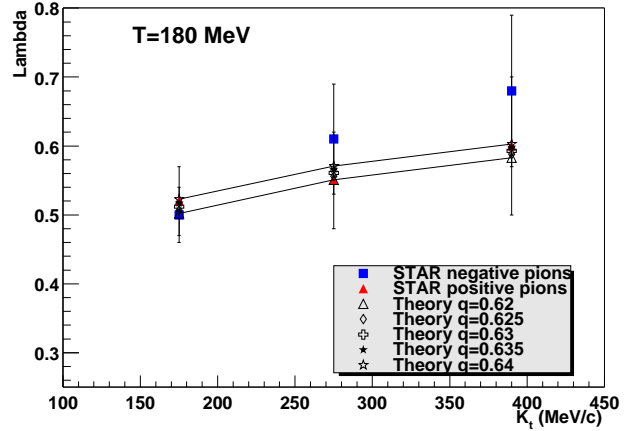}
%\includegraphics{fig1.eps}
%\vspace{-0.4cm}
\caption{ The transverse momentum $|{\bf K}_t|$
dependence of the intercept $\lambda $ of two-pion correlation.
Deformation parameter $q$ is a real quantity, $0 \le q \le 1$. }
  \label{Fig_1}
\vspace{-0.2cm}
\end{center}
\end{figure}

\begin{figure}
%[t]
\begin{center}
\vspace{-0.01cm}
\includegraphics[width=9cm]{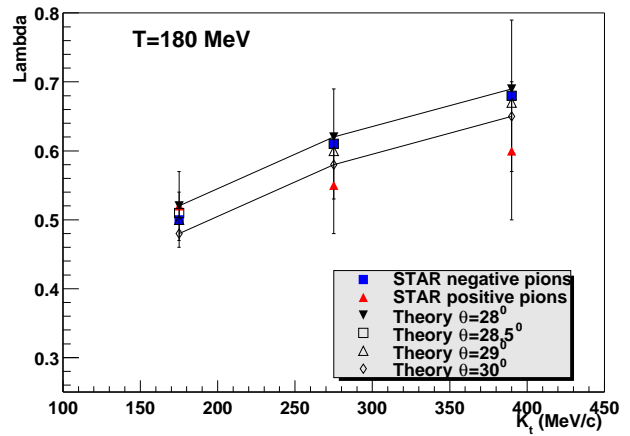}
%\includegraphics{fig2.eps}
%\vspace{-0.4cm}
\caption{The transverse momentum $|{\bf K}_t|$ dependence of the
intercept $\lambda $ of two-pion correlation. The deformation
parameter $q$ is taken in the form $q=e^{i \theta}$.}
\vspace{-0.2cm}
\label{Fig_2}
\end{center}
\end{figure}

Below, the two versions (\ref{10}) and (\ref{12}) corresponding to
BM- and AC-types of $q$-deformation are compared to the recent
STAR/RHIC data.
The experimental values for intercept parameter $\lambda$ in
Figs.~\ref{Fig_1} and~\ref{Fig_2} are taken from Ref.~\cite{star}.
The theoretical values are obtained by averaging over given
rapidity y and transverse momentum $K_t$ intervals
$\Delta_j\equiv K_t^{j,{\rm max}} - K_t^{j,{\rm min}}$,
$j=1,2,3$:
\beq
\lambda_j=                                            \label{13}
\frac{1}{\Delta_y}\int\limits_{-\Delta_y/2}^{\Delta_y/2}{\rm d}
y \, \frac{1}{\Delta_j} \int\limits_{K_{t}^{j,{\rm min}}}
^{K_{t}^{j,{\rm max}}} {\rm d}K_{t}~\lambda(q,m,T,y,K_t),
\eeq
where $m$ is particle mass.

Expressions (\ref{12}) and (\ref{10}) for $\lambda(q,m,T,y,K_t)$
were used in (\ref{13}) for  obtaining theoretical points shown in
Fig.~\ref{Fig_1} and Fig.~\ref{Fig_2} respectively.
%%%%%%%%%%%%%%%%%%%%%%%%%%%%%%%% fig1 %%%%%%%%%%%%%%%%%%%%%%%%%%%%
%%%%%%%%%%%%%%%%%%%%%%%%%%%%%%%%%%%%%%%%%%%%%%%%%%%%%%%%%%%%%%%
%\vspace{-1.2cm}
%%%%%%%%%%%%%%%%%%%%%%%%%%%%%%%% fig2 %%%%%%%%%%%%%%%%%%%%%%%%%%%%
%%%%%%%%%%%%%%%%%%%%%%%%%%%%%%%%%%%%%%%%%%%%%%%%%%%%%%%%%%%%%%%
\noindent
One can see from these figures that the agreement of
experimentally measured values of intercept parameter $\lambda$
with the theoretically calculated ones is very good.

Detailed comparison with the experiment \cite{star} gives:
the values $\lambda_i$ obtained from (13) at real $q$, see (12),
fit better the three experimental values for the intercept
of $\pi^+\pi^+$ correlations.
On the other hand, the values calculated by (13) with $q$ a pure
phase factor, see (10), agree better with the three experimental
values for the intercept of $\pi^-\pi^-$ correlations.
Possible explanation of the observed difference between
experimental values of intercept for $\pi^-\pi^-$-pairs and
$\pi^+\pi^+$-pairs could be the influence of the Coulomb FSI
of these charged pions with the positive charge of fireball
protons.
Since the AGI-model predicts that parameter $\lambda$ will
asymptotically reach a constant value  $\lambda^{\rm asymp.}<1$,
determined by $q$ only, at sufficiently large
($500\!-\!600$~MeV/c) pion pair mean momentum $\vert{\bf K}\vert$,
in order to check this prediction measurements at higher $K_t$
are necessary.
Such measurements should be available in near future at RHIC.

\begin{figure}
%[htb]
\begin{center}
\includegraphics[width=9cm]{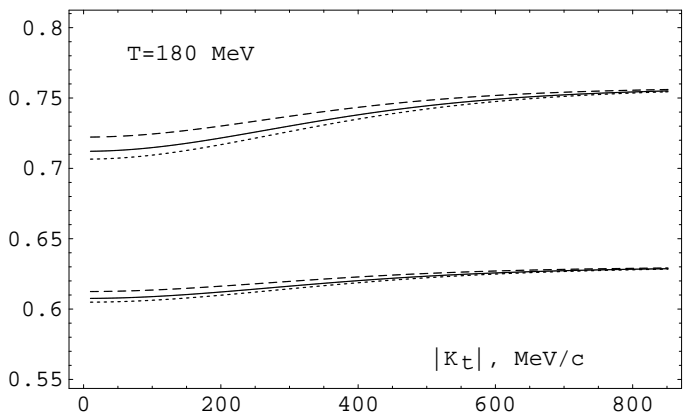}
\vspace{0.1cm}
\caption{
The transverse momentum $|{\bf K}_t|$ dependence of the intercept
$\lambda $ of two-kaon correlation.
Both the case of real $q=0.63$ (lower triple of curves) and the case
of $q=e^{i \theta}$ (upper triple of curves) with $\theta=28.5^\circ$
are shown. }
\vspace{-0.2cm}
\label{Fig_3}
\end{center}
\end{figure}

For the prediction of intercept of kaons, we will use the values
of $q$ which provide the best fit of experimental data for pions
(see Figs.~\ref{Fig_1} and~\ref{Fig_2}): $q=0.63$ or
$\theta=28.5^\circ$, assuming a universality of the deformation
parameter for description of excited hot hadronic matter.
The result of averaging in rapidity $-0.5\le y \le 0.5$, given by
first integral in (\ref{13}), is shown in Fig.~\ref{Fig_3} as
solid curve in each of the triples of curves.
The other two curves in each triple correspond to fixed
value of rapidity: $y=0$ (dotted curve) and $y=0.5$ (dashed curve).
Note that $y=0$ curve  and solid curve almost coincide.
%
%%%%%%%%%%%%%%%%%%%%%%%%%%%%%%%% fig3 %%%%%%%%%%%%%%%%%%%%%%%%%%%%
%%%%%%%%%%%%%%%%%%%%%%%%%%%%%%%%%%%%%%%%%%%%%%%%%%%%%%%%%%%%%%%
%
\noindent
As it is clearly seen, the cases of real $q$ and $q$ a phase
factor supply significantly different values for the kaon
intercept $\lambda_K$.
It is tempting to use just this feature for
making preference of particular version of the deformation
parameter $q$ - real or pure phase.
The choice is important because differing physics is behind these two
versions: real $q$ may reflect for instance particle finite size effects
\cite{anch92} or particle composite structure  \cite{avancini95},
and complex valued $q$ may refer to deformed symmetrization
properties of wave functions (like in Aharonov-Bohm effect),
relevant for short-lived systems occurred in heavy-ion collisions.
It is also possible that a phase-type $q$ encodes   \cite{gavr}
the effects from mixing at the composite (quark) level.
Recent data from NA44  \cite{NA44}, i.e. $\lambda=0.84\pm 0.13$
and $\lambda=0.61\pm 0.36$ resp. for
$\langle K_t \rangle \approx 0.25$~GeV/c and
$\langle K_t\rangle\approx 0.91$~GeV/c do not yet help in making
choice of optimal version for $q$.

In summary, we have presented comparison of the AGI model with
experimental data on two-particle correlations at RHIC, and found
remarkable agreement.
We used the parameters extracted from comparison with pion's data
to predict behavior of intercept of kaon correlation functions.
We stress again the crucial importance of correlation measurements
at high transverse momenta in order {\em to check the predicted
asymptotical "saturation" of intercept parameters}.
Measurements $\vert{\bf K}\vert$ in the range up to $500\!-\!600$
MeV/c for pions (up to $700\!-\!800$ MeV/c for kaons) should be
possible by RHIC detectors such as STAR and PHENIX.
The asymptotical behavior of intercept parameter $\lambda$ within the
proposed model, see (\ref{11}) for phase-type $q$, should
determine the actual value of the deformation parameter $q$
supposed to be a universal quantity for relativistic heavy ion
collisions.

%%%%%%%%%%%%%%%%%%%%%%%%%%%%%%%%%%%%%%%%%%%%%%%%%%%%%%%%%%%%%%%%
%%\section*{Acknowledgment}
%%%%%%%%%%%%%%%%%%%%%%%%%%%%%%%%%%%%%%%%%%%%%%%%%%%%%%%%%%%%%%%%%

D.A. acknowledges stimulating discussions with U.~Heinz, valuable
advices from P.~Braun-Munzinger and expresses his gratitude to
L.~McLerran and Nuclear Theory Group (BNL) for fruitful
discussions and warm hospitality.
The work of A.G. was partially supported
by the Award No. UP1-2115 of the U.S. Civilian Research and
Development Foundation (CRDF).
%%%%%%%%%%%%%%%%%%%%%%%%%%%%% bibliography %%%%%%%%%%%%%%%%%%%%%%%%%%

\end{document}